\documentclass{mn2e}
\usepackage{graphicx}
 
\begin{document}
 
\title{Spectroscopy of planetary mass brown dwarfs in Orion}
\author[Lucas, Weights, Roche, Riddick]
{P.W.Lucas$^{1}$, D.J.Weights$^{1}$, P.F.Roche$^{2}$, F.C.Riddick$^{3}$\\
$^1$Dept. of Physical Sciences, University of Hertfordshire, College Lane,
Hatfield AL10 9AB, England.\\ email: pwl@star.herts.ac.uk\\
$^2$Astrophysics, Physics Dept., University of Oxford, 1 Keble Road, Oxford OX1 3RH, England.\\
$^3$Dept. of Astronomy \& Astrophysics, Penn State University, 525 Davey Lab,
University Park, PA 16802, USA.}
\maketitle
\begin{abstract}
We report the results of near infrared spectroscopy of 11 luminosity
selected candidate planetary mass objects (PMOs) in the Trapezium Cluster 
with Gemini South/GNIRS and Gemini North/NIRI. 6 have spectral types $\ge$M9,
in agreement with expectations for PMOs.
2 have slightly earlier types, and 3 are much earlier types which are
probably field stars. 4/6 sources with types $\ge$ M9 have pseudo-continuum 
profiles which confirm them as low gravity cluster members.
The gravity status of the other cool dwarfs is less clear but these remain candidate
PMOs. The derived number fraction of PMOs with M=3-15~M$_{Jup}$ is 1-14\%, 
these broad limits reflecting the uncertainty in source ages. However, the number
fraction with M$<20~$M$_{Jup}$ is at least $5\%$. These detections add 
significantly to the body of evidence that a planetary mass population
is produced by the star formation process.

\end{abstract}

\section{Introduction}

In recent years there have been several reported detections of free floating
brown dwarfs with planetary masses in very young galactic clusters (eg. 
Oasa, Tamura \& Sugitani 1999; Lucas \& Roche 2000; Zapatero Osorio et al.2000;
Allers et al.2006; Jayawardhana \& Ivanov 2006). 
Many of these candidates are not yet confirmed as cluster members with the
low temperatures predicted by theoretical models. However, at least 1 confirmed 
PMO probably has M$<10$~M$_{Jup}$: Cha 1109-7734 (Luhman et al.2005). Here we adopt 
a planetary mass threshold 
of 15~M$_{Jup}$ (0.014~M$_{\odot}$) in order to include sources on the
deuterium burning limit, which is at 13-14~M$_{Jup}$ for solar abundance 
(Chabrier et al.2000; Burrows et al.1997) but may be marginally higher
in Orion, which has a low D/H ratio (see O'Dell 2004 and references therein).
  
Lucas, Roche \& Tamura (2005) (hereafter LRT) detected 33 candidate PMOs in 
a deep imaging survey of the Trapezium Cluster, and showed that $<$13\% of 
cluster members have masses in the range 3-13~M$_{Jup}$. These masses were estimated
from source luminosities and an assumed age of 1~Myr, 
using the isochrone of the Lyon group.
Here we present spectra of a sample of candidates
chosen from LRT and Lucas \& Roche (2000) in an attempt to constrain the 
inital mass function (IMF) at planetary masses.
In Section 2 the observations are described. In Section 3 we determine
spectral types and surface gravity status and assess cluster membership.
We then compare the results with the predictions of pre-main sequence
models and discuss the IMF.

\section{Observations}

Near infrared spectra of PMO candidates in Orion
were obtained in two observing programmes: one with GNIRS at Gemini
South Observatory and one in classical mode with NIRI at Gemini North Observatory. 
The GNIRS data were taken in queue mode 
between November 2004 and November 2005. The NIRI data were obtained in the 
classical observing mode on 5-8 December 2003.

Useful results were obtained for 11 candidate PMOs (see Table 1).
All are located in the outer parts of the Trapezium Cluster where
the pervasive nebulosity is relatively faint. 
They were selected for low extinction ($A_V \la 5$ mag) which
reduces the chance of observing background stars (see LRT).
The GNIRS sample consisted of 7 candidate PMOs from LRT. 
With NIRI we observed 5 candidate PMOs, 
4 selected from LRT and 1 (022-115) from Lucas \& Roche (2000). 
For one of these 5 candidates,
188-658, the data were of poor quality, so it was reobserved with GNIRS. 
The GNIRS
data were taken with a short slit, so only one source could be observed
at a time. Slit orientations which minimise the local nebulosity gradient in 
were chosen. The long slit of the NIRI spectrograph was oriented so as to include 
2 or 3 sources simultaneously. Hence, an additional selection criterion, for the NIRI 
sources only, was a very weak local nebulosity gradient, in order to minimise the 
effects of the structured background on the extracted spectra.
In both programmes the background emission was subtracted using a 
3\arcsec nod between 2 positions along the slit at intervals of 2-4 minutes.
The GNIRS data were taken in cross dispersed mode with the 32 lines/mm
grating.
Here we focus on the data in 2 orders covering the H (1.65~$\mu$m) and K (2.2~$\mu$m) 
bands, where the data quality is best.
Slit widths of 0.\arcsec45 or 0.\arcsec675 were used, depending on the
seeing conditions. The spectral resolution was 
R=$\lambda/ \Delta \lambda \approx 950$ in the H and K bands with the 
0.\arcsec675 slit and R$\approx$1400 with the 0.\arcsec45 slit.
The NIRI spectra were taken with a 0.\arcsec75 slit 
and the f/6 camera. 
The K band grism was used for most observations but the H band grism was used
to take some data at relatively high airmass. 
The resolution was R$ \approx 520$ for all the NIRI data.

The data were reduced with the IRAF software package. The OH sky lines
were used to wavelength calibrate the GNIRS spectra, and argon lamp 
spectra were used for the NIRI data. F and G-type stellar standards 
were used to calibrate the 
throughput. 
The extracted spectra were cleaned of narrow noise spikes 
attributable to bad pixels, cosmic rays and time variable telluric OH emission
lines, using independent subsets of the data to ensure that genuine features
were not discarded.
Residual background features caused by
3 to 5 of the brightest HI, H$_2$ or HeI lines (see Marconi et al. 
1998) were also fitted and subtracted.
The independent subsets were then coadded and dereddened with the DEREDDEN
task in IRAF to produce the final results.

\begin{table}
\begin{center}
{\bf Table 1 - Source List and Integration Times}\\
\begin{tabular}{lcccccc}

Source$^*$ & K & H & J & A$_V$ & Data & Time\\
                &   &   &   &      & Type      & (min) \\
                &   &   &   &      &            & H,K    \\ \hline
152-717 & 17.61 &  18.39 &  19.38 & 3.7 & GNIRS & 120,120  \\
188-658 & 17.87 &  18.34 &  19.12 & 0.8 & GNIRS & 132,132  \\
137-532 & 17.20 &  18.27 &  19.39 & 5.3 & GNIRS & 120,120  \\
092-532 & 17.71 &  18.22 &  19.09 & 2.4 & GNIRS &  56,56  \\
107-453 & 17.94 &  18.71 &  19.46 & 0.3 & GNIRS & 252,252 \\
016-430 & 17.89 &  18.94 &  19.98 & 2.4 & GNIRS &  72,72 \\
057-247 & 17.90 &  18.86 &  19.92 & 2.8 & GNIRS & 96,96 \\
183-729 & 17.24 &  17.38 &  18.05 & 1.3 & NIRI & 32,124  \\
199-617 & 17.64 &  18.96 &  20.17 & 5.2 & NIRI & 56,180 \\
205-610 & 17.16 &  17.81 &  18.52 & 1.1 & NIRI & 56,180  \\
022-115 & -     &  18.41 & 19.28 & 2.1 & NIRI &   -,184  \\

\end{tabular}
\end{center}
\small $*$ Source names are coordinate based, following O'Dell \& Wong (1996).
022-115 was listed as 023-115 in Lucas et al.(2001) due to a small astrometric 
error.
\end{table}
\normalsize

\section{Results and Analysis}

\subsection{Spectral Typing}

In Fig.1 we present the GNIRS and NIRI H and K band spectra for the 
8 sources with strong water vapour absorption.
Fig.1 also includes a K band spectrum of 022-115, 1 of the 3 sources 
with little or no steam absorption, and
2 high quality spectra from the template samples described below.
Steam absorption causes broad depresssions
on either side of the peaks near 1.675~$\mu$m in the H band and
2.20~$\mu$m in the K band. In contrast 022-115 shows a smooth
Rayleigh-Jeans continuum, which is typical of normal
field stars with spectral type $\le$M3, eg. Leggett et al.(1996).
Cubic spline fits to the plotted spectra generated by
the IRAF task SPLOT are overplotted for the sources in Orion.
These fits (5th order for H band and 6th order for K band)
represent the pseudo-continuum quite well, without following the narrower 
structures in the data, which are simply due to noise.

In Table 2 we provide the spectral types derived from four separate steam absorption
indices, which were applied to the pseudo-continuum fits. The indices were calibrated 
using a template sample of 38 young M dwarfs for which 
high quality 1-2.5~$\mu$m spectra taken with the SpeX instrument on the NASA Infrared 
Telescope Facility were provided by K.Luhman (see Luhman 2006; Muench et al., in
prep). A second template sample,
composed of late M and L type field dwarfs with high quality infrared spectra from
the same instrument, was also provided by K.Luhman.   
The spectral types of the young M dwarfs were defined at optical wavelengths using the 
dwarf/giant average system of Luhman (1999). All are located in the Chamaeleon or Taurus star 
formation regions, which have similar ages to the Trapezium Cluster.
The indices are defined by the following flux ratios: 
WH=(F$_{\lambda}$(1.525))/F$_{\lambda}$(1.675).  
QH=F$_{\lambda}$(1.57)/F$_{\lambda}$(1.675)(F$_{\lambda}$(1.77)/F$_{\lambda}$(1.675))$^{1.282}$.
WK=F$_{\lambda}$(2.05)/F$_{\lambda}$(2.23).\\
QK=(F$_{\lambda}$(2.05)/F$_{\lambda}$(2.2))(F$_{\lambda}$(2.34)/F$_{\lambda}$(2.2))$^{1.262}$. 
The median F$_{\lambda}$ values within 0.02~$\mu$m intervals were used, eg. 
1.515-1.535~$\mu$m for F$_{\lambda}$(1.525). 
Relations between spectral type and index were fitted as cubic polynomials
for all 4 indices. These have a typical scatter of 0.5 subtypes.
The properties of the indices are fully described by Weights (2006) and 
Weights et al.(in prep). To summarize, the QH and QK indices are reddening independent 
measures of the water absorption on both sides of the H and K band flux peaks respectively.
Similar indices were used by Wilking, Greene \& Meyer (1999) and 
Lucas et al.(2001). The WH and WK indices measure the steam absorption only on the
short side of the H and K bands respectively and require accurate A$_V$ values in Table 1.
However, they provide an independent check on 
the results and WK is less affected than QK by any veiling due to hot 
circumstellar dust on the long side of the K bandpass.
\begin{figure*}
\begin{center}
\begin{picture}(200,360)
 
\put(0,0){\includegraphics{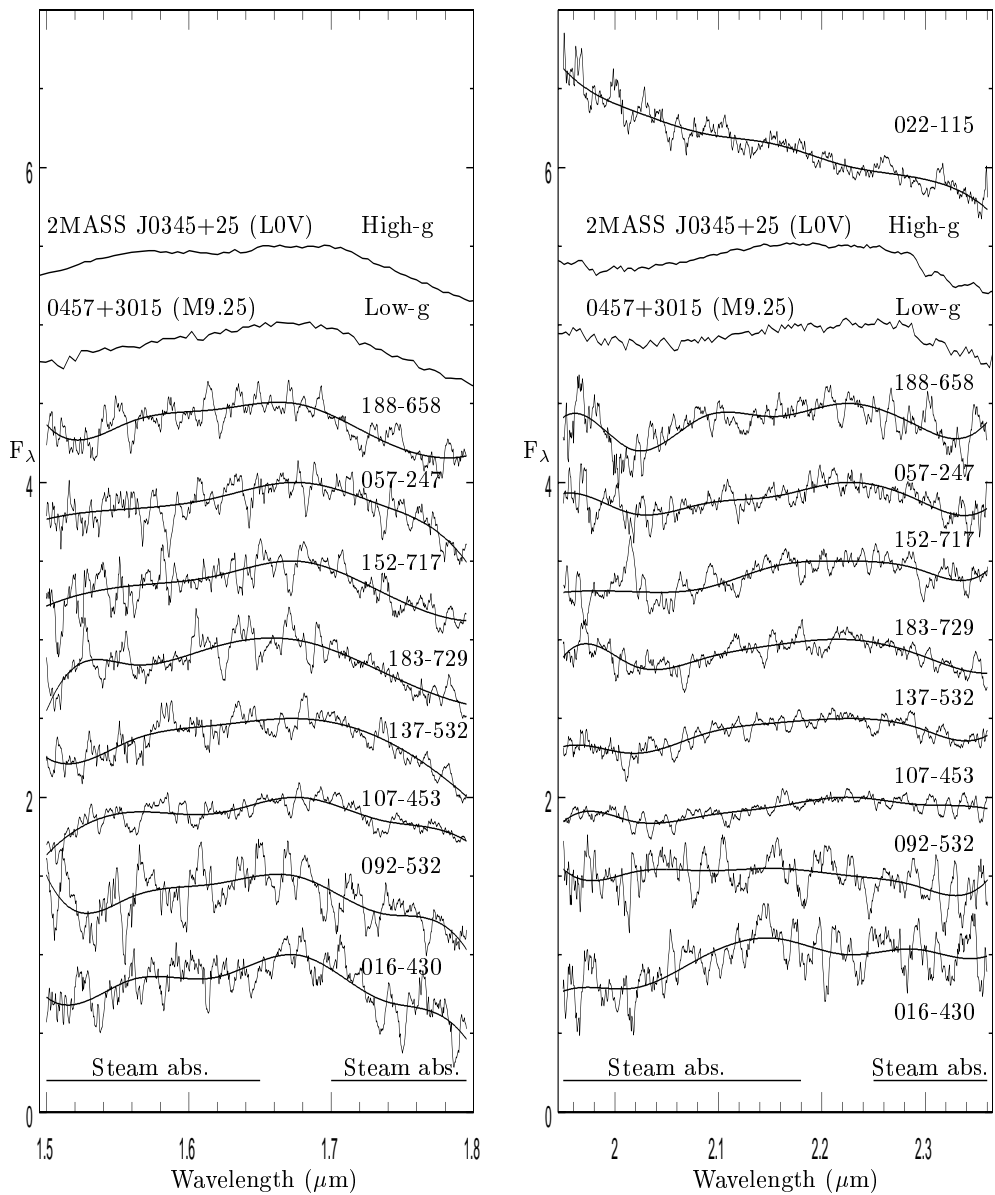}}

\end{picture}
\end{center}
\vspace{2.8cm}
\caption{\small Spectra of the 8 Trapezium sources with strong steam absorption bands
(the bottom 8 spectra) and 1 Trapezium source (022-115) with little or no steam absorption 
(top of right hand panel). An 11 pixel boxcar smoothing was applied. The fits to the 
pseudo continuum are overplotted as solid curves. 
Template spectra from the low-g and high-g samples provided by K.Luhman (see text) are 
also shown to illustrate the differences expected between low-g cluster 
members and cool high-g field dwarfs. Spectra are normalised to unity at the value of the 
curves at 1.675~$\mu$m and 2.23$\mu$m and are vertically displaced by multiples of 0.5 units 
for clarity.}
\end{figure*}
\normalsize

It should be noted that steam absorption strengthens rather slowly 
between M7 and M9 in the Luhman sample. The QK index discriminates best
between these types.
The quoted final types in Table 2 are based on comparison of the four indices 
and consideration of any source-specific aspects of the data (see notes to the 
Table). The final types are also influenced by the GNIRS J band spectra for the first 
7 objects in the table. The J band spectra (not shown) all exhibit a strong water 
absorption edge at 1.34~$\mu$m. Visual comparison with a large sample of field 
M dwarfs provided by S.Leggett (private comm.) shows that this is sufficient to 
classify these 7 objects as $>$M5 (see eg. Leggett et al.1996). We do not attempt 
to gain more precise information from the J band data, owing to the low signal to noise 
and the complicating effect of telluric water absorption near 1.34~$\mu$m.

\begin{table*}
\begin{minipage}{180mm}
\hspace{2cm}{\bf Table 2 - Spectral Types, Gravity and Cluster Membership}\\
\begin{tabular}{lccccccccccc}

Source & \multicolumn{4}{c}{Spectral Type} & Final  & 
\multicolumn{4}{c}{Spectral Type : Minimum $\chi^2$} & Gravity & Status \\
        & WH  & QH  & WK  & QK   & Type & Low-g H&High-g H & Low-g K& High-g K& &\\ \hline
152-717 & 9.3  & 10.3 & 10.2 & 9.2  &  $>$M9 &     $>$M9:2.0 & L3:3.2 & $>$M9:5.7 & L0:8.7 &  Low & PMO \\
188-658 & 9.7  & 9.5  & 10.7 & 10.5 &  $>$M9 &        M8:5.9 & L1:6.7 & M9:4.4 & L0:5.2 & Probably Low & 
Probable PMO \\
137-532 & 10.9 & 9.2  & 10.0 & 9.2  &  $>$M9 &     $>$M9:1.5 & L2:1.6 & $>$M9:5.0 & L0:9.8 &  Low & PMO \\
092-532 & 9.8  & 7.5  &  5.0 & 3.9  &  M7.5$\pm2^1$ & M6:2.9 & M7:3.3 & M6:2.9 & M7:3.2 & Uncertain & 
Uncertain \\
107-453 & 8.6  & 4.7  &  9.6 & 5.6  &  M8$\pm2^2$   & M9:4.1 & L1:3.8 & $>$M9:14.3 & L0:32.1 & Uncertain & 
Uncertain\\
016-430 & 11.5 & 10.7 &  9.3 & 6.4  &  M9$\pm2^3$   & M9:2.9 & L3:2.9 & M9:4.3 & L1:4.2 & Uncertain & Uncertain \\
057-247 & 8.8  & 9.3  & 10.1 & 10.0 &  $\ge$M9 &   $>$M9:2.0 & L1:3.2 & M8:1.9 & M9:3.1 & Low & PMO \\
183-729 & 7.6  & 10.6 & 10.0 & 10.2 &  $\ge$M9$^4$ &  M8:2.9 & L1:3.2 & M8:3.7 & M9:5.7 & Low$^4$ & PMO \\

\end{tabular}
\small Source notes: (1) greater weight was given to the H band indices for 
092-532, owing to the poorer quality of the K band spectrum. (2) The QK value was neglected, 
since 107-453 appears to have a K band excess, judging by the JHK fluxes and the K band 
continuum profile. (3) 016-430 shows a significant discrepancy between the WK and QK
results, probably due to the low signal to noise of the data. (4) The K band results were 
given more weight than H band for 183-729, due to the longer integration time. 183-729
was found to have low gravity and type M8.75$\pm 1.5$ in the optical spectrum of Riddick
(2006).
\end{minipage}
\end{table*} 
\normalsize

\subsection{Surface Gravity and Cluster Membership}

PMOs with the canonical 1~Myr age of Trapezium cluster members are predicted to have 
surface gravities 50 to 100 times lower than field dwarfs of similar spectral type, 
since they have not yet contracted to the stable radius (R$\approx$0.1~R$_{\odot}$) 
that is supported by degeneracy pressure in mature field dwarfs. The clearest 
indicator of cluster membership in very young 
sources is the detection of low surface gravity (low-g) features in the optical or
infrared spectrum (eg. McGovern et al. 2004). The most easily detected 
low-g feature is the triangular profile of the H band pseudo-continuum, which
peaks at $\sim 1.675~\mu$m in an F$_{\lambda}$ plot, identified by Lucas et al.(2001);
see also Natta et al.(2002), Meeus \& McCaughrean (2005). This contrasts with the flat 
topped profile 
observed in late M and L type high gravity (high-g) field dwarfs, a difference 
attributed by Kirkpatrick et al.(2006) to less collision induced H$_2$ absorption in 
low-g objects. In addition, the following systematic differences 
between the pseudo continua of low-g and high-g objects are observed in the K band in
the template samples. 
(1) F$_{\lambda}$ spectra of very young late M type brown dwarfs have a typically quite
flat maximum between 2.18 and 2.28~$\mu$m (eg. Luhman, Peterson \& Megeath 2004)
whereas field dwarfs with 
types M7-L7 all have a maximum between 2.14-2.18~$\mu$m and decline slowly from 
$\lambda = 2.18-2.29~\mu$m. This decline is more pronounced at mid-late L types.
(2) The CO (v=2-0) absorption trough located 
at $\lambda \ge 2.29~\mu$m is weaker in very young brown dwarfs with types
M7-M9.5 than in the field dwarfs with equivalent (optically determined) spectral type.

The surface gravity status of the 8 cool Orion dwarfs is indicated in Table 2. For 
some sources, eg. 152-717 and 057-247, the triangular H band profile is obvious from visual 
inspection. However, for several candidates the low signal to noise required a 
statistical test to determine whether the H and K band spectra data are more consistent 
with low-g template spectra than high-g field dwarf templates.
For each candidate reduced $\chi^2$ values were calculated separately
for the smoothed H and K band spectra using young dwarf and field dwarf
spectra from the template samples. Best fits were identified by
$\chi^2$ minimisation for a range of spectral types: M4 to M9.5
for the young dwarf templates and M4 to L3 for the field dwarf templates.
The uncertainties in F$_{\lambda}$ at each wavelength were estimated
using the scatter calculated in narrow wavelength bins and were
propagated through to the smoothed spectra. This will underestimate
the errors if there are also systematic effects, e.g. correction for atmospheric 
absorption. Hence, the absolute values of $\chi^2$ are difficult to interpret 
and are not used to derive formal probabilities. However, relative values can 
be used to determine whether high-g or low-g templates provide the better fits.

The $\chi^2$ results show that 4 sources have spectra
that are more consistent with low-g late M or early L type templates
than high-g templates in the H and/or K bands: 057-247, 152-717,
137-532, and 183-729 (Table 2). Low-g status is also probable for 188-658, 
though the result is less clear. For 107-453, the K-band fits are
poor owing to an unusually high flux at $\lambda > 2.2\mu$m. This may be
due to circumstellar dust emission, which would suggest that it is a 
young object.
The $\chi^2$ values for the best fits to 016-430 and 092-532 are not
significantly different for the low- and high-gravity templates.
The 3 sources exluded from Table 2 are almost certainly field stars, with spectral types 
$<$M5.

It is instructive to calculate the expected number of high-g field dwarfs within the 
26~arcmin$^2$ 
area of the LRT survey with similar spectral types and fluxes to the PMO candidates. 
The typical absolute magnitudes for single M9 and L5 field dwarfs are M$_K$=10.29, 11.61
respectively (Kirkpatrick et al. 2000) on the MKO system. Hence M9 and L5 field
dwarfs would appear fainter than all the sources in Table 1 at distances,
$d>$339~pc (M9) or 185~pc (L5). Spectral types $>$L5 were ruled out by the $\chi^2$ fits.
M9-L5 type field dwarfs have effective temperatures $1700<T_{eff}<2400$~K 
(Golimowski et al.2004) and a space density in the 
range 0.0067-0.0145~pc$^{-3}$ (Burgasser 2004). Multiplying the space density for 
each sub-type by the appropriate volume and summing the results, an expected contamination
of 0.36-0.72 M9-L5V single systems was calculated. This 
rises to 0.61-1.22 objects when unresolved binaries are included, assuming an incidence of 
25\% (Basri \& Reiners 2006) and equal luminosity for both components. If we 
conservatively assume a Poisson distribution with a mean of 1.22 and neglect candidates 
which have not yet been observed, the probability that all 6 
sources in Table 2 with types $\ge$M9 are field dwarfs is 0.2\% and the 
probability that more than 3 are field dwarfs is 3.6\%. 
When M6-M8 objects are included, the expected number of contaminating objects 
rises by a factor of $\sim 2$, so it is quite possible that 1 or 2 of the sources 
whose gravity status is uncertain are field dwarfs. 

\subsection{Comparison with model isochrones}
\begin{figure}
\begin{center}
\begin{picture}(100,170)
\put(0,0){\includegraphics{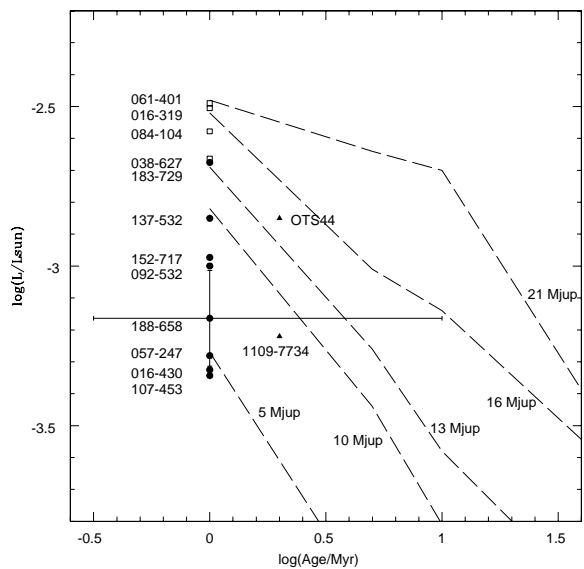}}
\end{picture}
\end{center}
\vspace{1.2cm}
\caption{\small Luminosity vs. age. The sources in Table 2 are plotted, with assumed ages of 
1~Myr (filled circles). The theoretical evolutionary tracks of
Chabrier et al.(2000) are plotted as dashed lines for various masses.
For comparison we plot
the 2 very low mass sources OTS44 and Cham 1109-7734 in the Chamaeleon star formation 
region (filled triangles) and the 4 least luminous Trapezium 
sources confirmed as cool dwarfs in Lucas et al.(2001) (open squares). Error bars are shown 
for only 1 source to aid clarity but are similar for all sources.}
\end{figure}
\normalsize

We estimate the masses of the cluster members 
by comparison with the model isochrones of Chabrier et al.(2000). The
masses can be derived from the isochrone using 
(i) luminosity and assumed age (see LRT), or (ii) temperature and assumed age.
In theory, both mass and age can be derived from luminosity and temperature
using the HR diagram but this method is too temperature sensitive at 2000-2500~K to 
produce useful results.
If an age of 1~Myr is assumed, 
method (i) gives masses M$\le 13$~M$_{Jup}$ in every case. 
Spectral types $\ge$M9 correspond to T$_eff \le 2400$~K, using the 
temperature scale of Luhman et al.(2003) that has been extensively applied in star
formation regions. At 1 Myr, method (ii) then indicates M$\le 15$~M$_{Jup}$. Hence the 
luminosities and temperatures agree with expectations for PMOs with an age 
of $\sim 1~$Myr. 
If an age of 10 Myr is assumed then the masses derived from method (i) increase somewhat, 
as shown in Fig.2. However, M$\le 20$~M$_{Jup}$ in every case, and we find
M$\le 15~M_{Jup}$ for the confirmed cluster member 057-257 and the likely cluster members 
188-658 and 016-430. For method (ii) a 
10~Myr age also gives M$\le 20$~M$_{Jup}$ for the 6 sources with spectral types $\ge$M9.
Since the cluster appears to contain fewer PMOs than more massive brown dwarfs
(see LRT) any age spread in the cluster would be expected to bias our 
luminosity selected sample towards larger ages than the cluster average. 
Most researches have shown that the great majority of stellar and substellar
cluster members have ages less than a few Myr
(Riddick 2006; Lucas et al.2001; Hillenbrand 1997; Lada et al.2004)
However, some studies suggest a broader age spread. Eg. Palla et al.(2005) detected lithium 
depletion in some low mass stars and proposed that star formation 
began in the cluster at a low rate $\sim10~$Myr ago and accelerated toward the present day. 


\subsection{The Initial Mass Function}

Since 10~Myr ages are possible for the PMO candidates, it is not yet proven
that any have mass M$< 10~M_{Jup}$. However, even at age 10~Myr at least 1 source
is confirmed with M$< 15~M_{Jup}$, along with 2 other likely cluster members.
The model isochrones also become highly uncertain at ages less than 
a few Myr (Baraffe et al.2002). However recent dynamical mass measurements of an eclipsing 
binary brown dwarf in the Trapezium cluster agree well with the predictions of the 
1~Myr isochrone (Stassun, Mathieu \& Valenti 2006) and less direct mass estimates based on 
surface gravities (Mohanty et al.2004) are also fairly consistent with the predictions 
for very young brown dwarfs.
Hence, at present the evidence suggests the model isochrones yield approximately correct
masses at ages as low as 1~Myr.

This dataset therefore adds significantly to the evidence that free floating PMOs
exist in very young clusters, extending the results of Lucas et al.(2001) to lower
luminosities. In a sample of 11 PMO candidates we have 4 confirmed
cluster members, 1 very probable member, 3 sources with uncertain status, and 3 
field stars. Assuming an age of 1~Myr, 
LRT found that PMO candidates with M=3-13~M$_{Jup}$ number 13.2\% of the 
population. Considering only the 7 sources whose status is clear, and 
assuming a 1~Myr age and Poisson statistics, a number fraction of confirmed 
PMOs of $7.5\% \pm 2.7\%$ 
is simplistically derived. However, $\sim$50\% of the candidates of 
LRT lay in the 3-5~M$_{Jup}$ mass bin, which is under-represented in this sample 
(see Fig.2) but may contain a higher fraction of field stars.
In addition, since the bias towards ages $>1~$Myr at very low luminosities
is not quantified, the number fraction of PMOs can only be loosely constrained. 
However, we can say that the PMO number fraction in the 
range 3-15~M$_{Jup}$ is almost certainly between 1\% and 14\%, and the fraction with 
M$<20~$M$_{Jup}$ is $>$5\%. 

\section{Acknowledgements}
\small
We thank Kevin Luhman for supplying the template spectra used to calibrate the
Orion spectra. This paper is based on observations obtained in programmes 
GS-2004B-Q-11 and GN-2003B-C-1 at the Gemini
Observatory, which is operated by the Association of Universities for Research
is Astronomy, Inc., under a cooperative agreement with the NSF on behalf
of the Gemini partnership: the National Science Foundation (USA), the
Particle Physics and Astronomy Research Council (UK), the National Research
Council (Canada), CONICYT (Chile), the Australian Research Council (Australia),
CNPq (Brazil) and CONICET (Argentina).  

\section{References}
\small

Baraffe I., Chabrier G., Allard F., Hauschildt P.H., 2002, A\&A, 382, 563\\
Basri G., Reiners A., 2006, AJ 132, 663\\
Burgasser A. 2004, ApJS, 155, 191\\
Chabrier G., Baraffe I., Allard F., Hauschildt P. 2000, ApJ 542, 464\\
Golimowski D.A., Leggett S.K., Marley M.S., Fan X., Geballe T.R., Knapp G.R., Vrba F.J., 
    Henden A.A., Luginbuhl C.B., Guetter H. H. and 9 coauthors, AJ 2004, 127, 3516\\
Hillenbrand L.A. 1997, AJ 113,1733\\
Jayawardhana R., \& Ivanov V.D., 2006, ApJ 647, L167 \\
Kirkpatrick J.D., Reid I.N., Liebert J., Gizis J.E., Burgasser A.J., Monet D.G., 
    Dahn C.C., Nelson B., Williams R,J., 2000, AJ 120, 447\\
Kirkpatrick J.D., Barman T.S., Burgasser A.J., McGovern M.R., McClean I.S.,
    Tinney C.J., Lowrance P.J. 2006, ApJ, 639, 1120\\
Lada C.J., Muench A.A., Lada E.A., Alves J.F. 2004, AJ, 128, 1254\\
Leggett S.K., Allard F., Berriman G., Dahn C.C., 1996, ApJS 104, 117\\
Lucas P.W., Roche P.F., 2000, MNRAS 314, 858\\
Lucas P.W., Roche P.F., Allard F., Hauschildt P.H. 2001, MNRAS 326, 695\\
Lucas P.W., Roche P.F., Tamura M., 2005, MNRAS 361, 211\\
Luhman K.L., 1999, ApJ, 525, 466 \\
Luhman K.L., Stauffer J.R., Muench A.A., Rieke G.H., Lada E.A., Bouvier J., Lada C.J., 
    2003, ApJ 593, 1093\\
Luhman K.L., Peterson D.E., Megeath S.T. 2004, ApJ 617, 565\\
Luhman K.L., Adame L., D'Alessio P., Calvet N., Hartmann L., Megeath S.T.,
    Fazio G.G., 2005, ApJ 635, L93\\
Luhman K.L., 2006, ApJ 645, 676\\
McGovern M.R., Kirkpatrick J.D., McLean I.S., Burgasser A.J., Prato L., 
    Lowrance P.J., 2004, ApJ 600, 1020\\
Meeus G., McCaughrean M.J., 2005, AN 326, 977\\
Mohanty S., Jayawardhana R., Basri G. 2004, ApJ 609, 885\\
Natta A., Testi L., Comeron F., Oliva E., D'Antona F.D., Baffa C., 
    Comoretto G., Gennari S., 2002, A\&A 393, 597\\
Oasa Y., Tamura M., \& Sugitani K. 1999, ApJ 526,336\\
O'Dell C.R., 2001, ARA\&A 39, 99\\
O'Dell C.R., Wong K., 1996, AJ 111, 846\\
Palla F., Randich S., Flaccomio E., Palla R., 2005, ApJ 626, L49\\
Riddick F.C., 2006, PhD thesis, Oxford University.\\
Stassun K.G., Mathieu R.D., Valenti J.A., 2006, Nature 440, 311\\
Weights D.J., 2006, http://star-www.herts.ac.uk/$\sim$dweights\\
Zapatero Osorio M.R., B\'{e}jar, V.J.S., Mart\'{i}n E.L., Rebolo R., Barrado y 
    Navascu\'{e}s D., Bailer-Jones C.A.L., Mundt R., Science 290, 103\\

\end{document}